\documentclass[
  prd,
  aps,
  showpacs,
  eqsecnum,
  notitlepage,
  nofootinbib,
  superscriptaddress,   
  colorlinks=true,
  linkcolor=blue,
  citecolor=blue,
  urlcolor=blue,
  pdfstartview=FitV,
  breaklinks=true,
  10pt,
  twocolumn
]{revtex4-2}

\usepackage[utf8]{inputenc}
\usepackage[T1]{fontenc}
\usepackage{lmodern}

\usepackage{graphicx}
\usepackage{tikz}
\usepackage{siunitx}
\usepackage{tensor}
\usepackage{amsmath,amssymb,amsfonts,bm}
\usepackage{slashed}
\usepackage{enumerate}
\usepackage{scalerel}       
\usepackage{accents}

\usepackage[caption=false]{subfig}

\usepackage{orcidlink}

\begin{document}

\title{Elastic field causing noncommutativity}
\author{A. L. Silva Netto\,\orcidlink{https://orcid.org/0000-0002-6988-5041}}
\email{anibal.livramento@univasf.edu.br}
\affiliation{Universidade Federal do Vale do S\~ao Francisco, 48902-300 Juazeiro, BA, Brasil.}

\author{A.~M.~de~M.~Carvalho\,\orcidlink{0009-0006-3540-0364}}
\email{alexandre@fis.ufal.br}
\affiliation{Instituto de F\'{\i}sica, Universidade Federal de Alagoas, 57072-970 Macei\'o, AL, Brazil}

\author{G.~Q.~Garcia\,\orcidlink{0000-0003-3562-0317}}
\email{gqgarcia99@gmail.com}
\affiliation{Centro de Ci\^encias, Tecnologia e Sa\'ude, Universidade Estadual da Para\'iba, 58233-000 Araruna, PB, Brazil}

\author{C.~Furtado\,\orcidlink{0000-0002-3455-4285}}
\email{furtado@fisica.ufpb.br}
\affiliation{Departamento de F\'isica, Universidade Federal da Para\'iba, 58051-970 Jo\~ao Pessoa, PB, Brazil}


\begin{abstract}
We study how a uniform torsion background—modeling a continuous density of screw dislocations—induces effective spatial noncommutativity and reshapes the energy spectrum of a free quantum particle. Within the geometric theory of defects, the metric yields a first-order (magnetic-like) coupling in the transverse dynamics, equivalent to an effective magnetic field $B_{\mathrm{eff}}\propto p_z\,\Omega$, where $\Omega$ encodes the torsion strength. In the strong-coupling (Landau) regime, the planar coordinates obey $[x,y]\neq 0$ and the spectrum organizes into Landau-like levels with a slight electric-field-driven tilt and a uniform shift. Thus, increasing $\Omega$ drives the system continuously toward the familiar Landau problem in flat space, with torsion setting the noncommutativity scale and controlling the approach to the Landau limit.
\end{abstract}


\maketitle
%
\section{Introduction}
\label{intro}
The most familiar notion of noncommutativity in quantum mechanics originates in the Heisenberg uncertainty principle: specific pairs of operators representing physical observables do not commute, so one cannot measure both with arbitrarily high precision at the same time~\cite{heisenberg1927}.
In the 1930s, Heisenberg suggested in a letter to Peierls that even the spatial coordinates themselves might be noncommutative, as a possible way to tame ultraviolet divergences in quantum field theory; he did not yet have a clear physical interpretation of such coordinate uncertainty.
The idea circulated among Heisenberg, Pauli, and Oppenheimer, and was developed rigorously by Snyder, who constructed a Lorentz-invariant noncommutative spacetime~\cite{snyder1947}.
In parallel, Weyl and Moyal laid the foundations for noncommutative structures in phase-space quantum mechanics~\cite{weyl1931,moyal1949}.
With the later success of renormalization, interest waned for some time. Still, noncommutative geometry returned to the mainstream decades later through the work of Connes and collaborators, who endowed it with a robust differential-geometric framework~\cite{connes1994}.

Arguments from quantum gravity suggest the existence of a minimal length near the Planck scale \( \ell_{\mathrm P} \), rather than an experimentally established bound.
In general relativity, a related intuition appears in Thorne's hoop conjecture: sufficiently concentrated energy forms a horizon when confined within a region whose size is comparable to the Schwarzschild radius; attempts to localize an event below that scale would trigger gravitational collapse~\cite{kipthorne}.
On the quantum side, the Heisenberg uncertainty principle limits the simultaneous precision of conjugate observables.
In condensed matter, an electron in a uniform magnetic field perpendicular to the plane provides a concrete arena where effective coordinate noncommutativity emerges in the lowest Landau level, and related noncommutative models have been explored as probes of extensions of the Standard Model.
Examples include spontaneous symmetry breaking~\cite{spontaneousbreaking}, thermodynamic and statistical effects for bosons and fermions~\cite{thermodynamics}, geometric phases~\cite{geophase1,geophase2,geophase3,geophase4,geophase5}, and Landau-level physics~\cite{landau1,landau2}.

The operator algebra defines the noncommutative space considered in this work
\begin{equation}
\begin{split}
&[\hat{x}^l,\hat{x}^m]= i\,\theta^{lm}\,\mathbb{I},\\
&[\hat{p}^l,\hat{p}^m]= 0,\\
&[\hat{x}^l,\hat{p}^m]= i\hbar\,\delta^{lm}\,\mathbb{I},
\end{split}
\label{relatcommutation}
\end{equation}
where $\mathbb{I}$ is the identity and $\theta^{lm}$ is a constant antisymmetric tensor. In the usual (commutative) case, the first commutator in~\eqref{relatcommutation} vanishes. A direct consequence is that spatial localization becomes coarse-grained into finite cells with area set by $|\theta^{lm}|$, a feature often referred to as fuzzy space~\cite{madore}. And we connect these ideas to defects in elastic media, building on the classical works of Kr\"oner~\cite{kroner}, Nye~\cite{nye}, and Bilby~\cite{bilby}, the geometric theory of defects (GTD) models continua containing disclinations and dislocations within a Riemann--Cartan geometry~\cite{kleinert1,volovik,bausch,katanaev1}. Geometric objects such as metric, curvature, and torsion describe, respectively, elastic distances and the strengths of disclinations and dislocations; the corresponding intensities are encoded by the Frank angle and the Burgers vector~\cite{katanaev2}. In this framework, the metric becomes non-Euclidean in the presence of topological defects, and homogeneous defect distributions generate homogeneous geometric fields.

In this work, we use the geometric theory of defects (GTD) to model a medium with a uniform areal density of screw dislocations, which produces uniform torsion throughout space. We show that this torsion acts as an effective magnetic field and that the resulting background leads to noncommuting coordinates, in line with the noncommutative dynamics discussed in Refs.~\cite{magro,fadeev}. We determine the resulting noncommutativity scale in terms of a single geometric parameter $\Omega$, which is fixed directly by the dislocation density. The paper is organized as follows. Section~\ref{geoview} summarizes the geometric background for screw dislocations and derives the Hamiltonian from the metric. Section~\ref{noncommutsection} establishes the commutation relations for the coordinate operators and makes explicit their dependence on the torsion strength $\Omega$. Section~\ref{application} applies the formalism to the quantum dynamics in the presence of a uniform electric field and the effective magnetic field, highlighting the resulting Landau-like spectrum and its corrections. Finally, Section~\ref{conclusions} presents our concluding remarks.

\section{Geometric Theory of Defects}\label{geoview}
In this section, we collect the geometric ingredients for a medium containing screw dislocations. We begin with a single screw dislocation obtained via the classical Volterra process~\cite{volterra}. Consider an elastic medium with a cut along a half-plane, and perform a relative translation of the faces by an amount $b$ parallel to the dislocation line. The vector $\vec{b}$ is the Burgers vector of the defect. A standard line element for a screw dislocation aligned with the $z$ axis is
\begin{equation}
ds^{2} = \bigl(dz+\beta\,d\varphi\bigr)^{2} + d\rho^{2} + \rho^{2}\,d\varphi^{2},
\label{screwmetric}
\end{equation}
with Burgers magnitude related by $b=2\pi\beta$. For later use, we write the metric in an orthonormal coframe $e^a = e^{a}_{\ \mu}(x) dx^\mu$,
\begin{equation}
e^{1}=dz+\beta\,d\varphi,\qquad
e^{2}=d\rho,\qquad
e^{3}=\rho\,d\varphi,
\label{screw1form}
\end{equation}
where we can describe the metric tensor as $g=\delta_{ab}\,e^{a}\!\otimes e^{b}$. 

In a Riemann-Cartan background, the torsion two-forms are given by the Maurer-Cartan structures,
\begin{equation}
T^{a}=de^{a}+\omega^{a}{}_{b}\wedge e^{b},
\label{eq:structure}
\end{equation}
with a metric-compatible connection $\omega^{a}{}_{b}$. For a single screw dislocation, one may choose $\omega^{a}{}_{b}$ so that the geometry is flat ($R^{a}{}_{b}=0$) away from the core and the defect is encoded in a distributional torsion at $\rho=0$. Using \eqref{screw1form} and the identity $d \left(d\varphi\right) = 2\pi\,\delta^{(2)}(\rho)\,d\rho\wedge d\varphi$, we obtain
\begin{subequations}
\begin{eqnarray}
\label{1formscrew}
T^{1} &=& d(e^{1}) = 2\pi\beta\,\delta^{(2)}(\rho)\,d\rho\wedge d\varphi,\\
T^{2} &=& 0;\\ 
T^{3} &=& 0,
\end{eqnarray}
\end{subequations}
in other words, the torsion is localized at the dislocation line. The flux of torsion through a disk $\Sigma$ encircling the core gives the Burgers vector via Stokes:
\begin{equation}
\int_{\Sigma} T^{1}=\oint_{\partial\Sigma} e^{1}
=\oint_{\partial\Sigma} (dz+\beta\,d\varphi)
=2\pi\beta=b.
\label{fluxtorsion1}
\end{equation}
Equation~\eqref{fluxtorsion1} is the differential-form version of the Burgers circuit: the torsion flux equals the Burgers strength. By analogy with Gauss's law, the flux equals the source; here this follows from Stokes' theorem and the Cartan structure equations, and the simple flux form holds in regions where $R^{a}{}_{b}=0$.

We now consider a continuous, homogeneous distribution of parallel screw dislocations in an elastic medium. A convenient metric encoding this background is
\begin{equation}
ds^{2}=\bigl(dz+\Omega\,\rho^{2}d\varphi\bigr)^{2}+d\rho^{2}+\rho^{2}d\varphi^{2},
\label{screwmetric2}
\end{equation}
with coframe defined by
\begin{equation}
e^{1}=dz+\Omega\,\rho^{2}d\varphi,\qquad
e^{2}=d\rho,\qquad
e^{3}=\rho\,d\varphi .
\end{equation}
From $T^{a}=de^{a}+\omega^{a}{}_{b}\wedge e^{b}$ and $d(\rho^{2}d\varphi)=2\rho\,d\rho\wedge d\varphi$ one finds
\begin{subequations}
\begin{eqnarray}
T^{1} &=& d(e^{1}) = 2\Omega\,\rho\,d\rho\wedge d\varphi;\\
T^{2} &=& 0;\\
T^{3} &=& 0,
\end{eqnarray}
\end{subequations}
so that the torsion is uniform in the orthonormal frame, $T^{1}{}_{\rho\varphi}=2\Omega$. The torsion flux through a disk of radius $R$ reads
\begin{equation}
\int_{\Sigma}T^{1}=\int_{0}^{R}\!\!\int_{0}^{2\pi}2\Omega\,\rho\,d\rho\,d\varphi
=2\pi\,\Omega\,R^{2},
\label{fluxtorsion2}
\end{equation}
which matches the Burgers content of a homogeneous distribution. Identifying the areal density $n$ of screw dislocations (Burgers magnitude $b$) via $\int_{\Sigma}T^{1}=n\,b\,A(\Sigma)$ gives
\begin{equation}
T^{1}{}_{\rho\varphi}=n\,b \quad\Rightarrow\quad \Omega=\tfrac12\,n\,b .
\end{equation}
By the same Gauss-type analogy as before, the “flux equals source” interpretation follows from Stokes' theorem; here it simply quantifies the uniform dislocation content.

For a spinless particle in the background \eqref{screwmetric2}, and in the absence of an external electromagnetic field, the Hamiltonian is the Laplace--Beltrami operator
\begin{equation}
H=-\frac{\hbar^{2}}{2\mu}\,\frac{1}{\sqrt{g}}\,
\partial_{i}\!\left(\sqrt{g}\,g^{ij}\partial_{j}\right),
\label{HBare}
\end{equation}
with $i,j\in\{\rho,\varphi,z\}$. Here $\sqrt{g}=\rho$ and the nonvanishing inverse components are
$g^{\rho\rho}=1$, $g^{\varphi\varphi}=1/\rho^{2}$, $g^{zz}=1+\Omega^{2}\rho^{2}$, and
$g^{z\varphi}=g^{\varphi z}=-\Omega$; equivalently, in the covariant metric one has
$g_{\rho\rho}=1$, $g_{\varphi\varphi}=\rho^{2}$, $g_{zz}=1$, and the only off-diagonal entry is
$g_{z\varphi}=g_{\varphi z}=\Omega\rho^{2}$, which generates first-order terms that act as a magnetic-like coupling in the reduced planar dynamics. If a real electromagnetic field is present, we include minimal coupling as
\begin{equation}
H=-\frac{\hbar^{2}}{2\mu}\,\frac{1}{\sqrt{g}}\,
\left(\partial_{i}-\frac{ie}{\hbar c}A_{i}\right)
\sqrt{g}\,g^{ij}
\left(\partial_{j}-\frac{ie}{\hbar c}A_{j}\right),
\label{hamiltdef}
\end{equation}
but in what follows we set $A_{i}=0$ and attribute the magnetic-like effects solely to the geometry \eqref{screwmetric2}.

\section{Noncommutativity of Landau Levels}\label{noncommutsection}
We now probe the noncommutative character of the space under study. Our goal is to obtain the commutators between operators associated with spatial coordinates. As a warm-up, recall the standard Landau problem. In Cartesian coordinates, we choose the symmetric gauge
\begin{equation}
\vec{A}=\left(-\tfrac{B}{2}\,y,\;\tfrac{B}{2}\,x,\;0\right),
\label{gauge}
\end{equation}
where $B$ points along the $z$ direction. The Lagrangian for a particle of mass $\mu$ and charge $e$ in an electromagnetic field is
\begin{equation}
L=\frac{\mu}{2}\bigl(\dot{x}^{2}+\dot{y}^{2}+\dot{z}^{2}\bigr)+\frac{e}{c}\,\bigl(\dot{x}A_{x}+\dot{y}A_{y}+\dot{z}A_{z}\bigr),
\label{lagrangian}
\end{equation}
which yields the Hamiltonian
\begin{equation}
H=\frac{1}{2\mu}\left(\vec{p}-\frac{e}{c}\vec{A}\right)^{2},
\end{equation}
with Landau energies
\begin{equation}
E_{n}=\hbar\omega_{c}\left(n+\tfrac{1}{2}\right),\qquad
\omega_{c}=\frac{eB}{\mu c}.
\label{landauenergy}
\end{equation}
In the strong-field (or lowest-Landau-level) regime one can neglect the quadratic kinetic term in the plane and keep only the first-order  contribution,
\begin{equation}
L_{B}=\frac{eB}{2c}\,(x\dot{y}-y\dot{x}).
\label{strongfield}
\end{equation}
From \eqref{strongfield} it follows that the equal-time commutator of the planar coordinates is
\begin{equation}
[\hat{x},\hat{y}]= i\,\frac{\hbar c}{eB}\,\mathbb{I}.
\label{commut2}
\end{equation}

We now map the geometric background \eqref{screwmetric2} to the Landau form. 
Due to translational invariance along the $z$-direction, the canonical momentum 
$p_z$ is conserved. In what follows, we work within a sector of fixed longitudinal 
momentum, $p_{z} = \hbar k$, so that the three-dimensional dynamics reduces to a practicalThe  planar problem in the $(x,y)$ coordinates. In this sector, the Lagrangian 
in the plane acquires a first-order term
\[
    L_{\text{geom}} = p_{z}\,\Omega\,\rho^{2}\dot{\varphi} 
    \;=\; \frac{p_{z}\Omega}{2}\,(x\dot{y}-y\dot{x}),
\]
which has exactly the structure of \eqref{strongfield}. Identifying the coefficients 
in $L_{\rm geom}=\frac{p_z\Omega}{2}(x\dot y-y\dot x)$ and 
$L_B=\frac{eB}{2c}(x\dot y-y\dot x)$ then gives
\begin{equation}
    B_{\rm eff}=\frac{c}{e}\,p_z\,\Omega .
    \label{Beff-correct}
\end{equation}
Within each fixed-$p_z$ sector, the standard Landau commutation relation 
$[x,y]= i\,\hbar c/(eB)$ now applied to $B_{\rm eff}$ leads to the geometric 
noncommutativity
\begin{equation}
    [\hat x,\hat y]= i\,\frac{\hbar}{p_z\Omega}\,\mathbb{I},
    \label{xy-comm-correct}
\end{equation}
or, in polar coordinates
\begin{equation}
[\hat\rho,\hat\varphi]= i\,\frac{\hbar}{p_z\Omega}\,\frac{1}{\rho}\,\mathbb{I}.
\label{rphi-comm-correct}
\end{equation}
Equations~\eqref{xy-comm-correct}–\eqref{rphi-comm-correct} show that the
noncommutativity scale is inversely proportional to both the torsion strength
$\Omega$ and the conserved longitudinal momentum $p_{z}$: increasing torsion
or $p_{z}$ shrinks the noncommutative length scale associated with the
$(x,y)$ plane. The polar-coordinate commutator in Eq.~\eqref{rphi-comm-correct} should be understood for $\rho\neq 0$, i.e., away from the symmetry axis where $\varphi$ becomes ill-defined.

\section{Quantum dynamics under a uniform electric field and $B_{\text{eff}}$}
\label{application}
In previous work~\cite{anibal}, we showed that the space with metric \eqref{screwmetric2} plays the role of an effective magnetic field, and we found elastic Landau levels. We now study a medium subjected to both a uniform electric field and the effective magnetic field. We choose the symmetric gauge
\begin{equation}
\vec{A}=\left(-\frac{\Lambda}{2}\,y\,,\,\frac{\Lambda}{2}\,x\,,\,0\right),
\label{magdef}
\end{equation}
where $\Lambda$ denotes the effective magnetic field (cf. Sec.~\ref{noncommutsection}, $B_{\text{eff}}=\Lambda$), and take the electric potential
\begin{equation}
\phi=-\mathcal{E}x \quad \Rightarrow \quad \vec{\mathcal{E}}=-\nabla\phi=\mathcal{E}\,\hat{\mathbf x}.
\label{electricdef}
\end{equation}
The Hamiltonian is then
\begin{equation}
H(\vec r,\vec p)=\frac{1}{2\mu}\!\left[\Big(p_{x}+\frac{e\Lambda}{2c}\,y\Big)^{2}
+\Big(p_{y}-\frac{e\Lambda}{2c}\,x\Big)^{2}\right]
+e\mathcal{E}x.
\label{H_EB_eff}
\end{equation}
In order to solve the eigenvalue problem $\hat{H}\Psi=E\Psi$, with $\hat{H}=H(\hat{\vec r},\hat{\vec p})$, it is convenient to introduce complex coordinates and momenta
\begin{subequations}
\begin{eqnarray}
\hat{z} &=& \hat{x}+i\hat{y};\\
\hat{p}_{z} &=& \tfrac{1}{2}\bigl(\hat{p}_{x}-i\hat{p}_{y}\bigr);\\
\hat{p}_{\bar{z}} &=& \tfrac{1}{2}\bigl(\hat{p}_{x}+i\hat{p}_{y}\bigr),
\end{eqnarray}
\end{subequations}
with canonical quantization $[\hat r_{i},\hat p_{j}]=i\hbar\,\delta_{ij}\,\mathbb{I}$, so that
$[\hat z,\hat p_{z}]=i\hbar\,\mathbb{I}$ and $[\hat{\bar z},\hat p_{\bar z}]=i\hbar\,\mathbb{I}$.
We also define the effective cyclotron frequency
\begin{equation}
\nu \equiv \frac{e\Lambda}{\mu c}.
\label{nu_def}
\end{equation}

Introducing the formalism of creation/annihilation operators
\begin{subequations}
\begin{eqnarray}
\label{a_ops}
a^{\dagger} &=& -2i\,\hat{p}_{\bar z}+\frac{e\Lambda}{2c}\,\hat{z}+\lambda;\\
a &=&\,2i\,\hat{p}_{z}+\frac{e\Lambda}{2c}\,\hat{\bar z}+\lambda,
\end{eqnarray}
\end{subequations}
and
\begin{subequations}
\begin{eqnarray}
b^{\dagger} &=& -2i\,\hat{p}_{\bar z}-\frac{e\Lambda}{2c}\,\hat{z};\\
b &=&\,2i\,\hat{p}_{z}-\frac{e\Lambda}{2c}\,\hat{\bar z},
\label{b_ops}
\end{eqnarray}
\end{subequations}
where
\begin{equation}
\lambda=\frac{\mu c\,\mathcal{E}}{\Lambda}.
\label{lambda_def}
\end{equation}
With this normalization, $[a,a^{\dagger}]=2\mu\hbar\,\nu$ and $[b,b^{\dagger}]=2\mu\hbar\,\nu$, while $[a,b]=[a,b^{\dagger}]=0$. So, the Hamiltonian operator can be written as
\begin{equation}
\hat{H}=\hat{H}_{\text{osc}}-\hat{K},
\end{equation}
with
\begin{equation}
\hat{H}_{\text{osc}}=\frac{1}{4\mu}\bigl(a^{\dagger}a+aa^{\dagger}\bigr)
\,\ \text{and}\,\ 
\hat{K}=\frac{\lambda}{2\mu}\,(b^{\dagger}+b)+\frac{\lambda^{2}}{2\mu}.
\end{equation}
Thus, the problem separates into an oscillatory part $\hat{H}_{\text{osc}}$ and a linear part $\hat{K}$.

First of all, we deal with the eigenvalue equation for the oscillatory part $\hat{H}_{\text{osc}}$. Therefore, we have the Schr\"odinger equation
\begin{equation}
\hat{H}_{\text{osc}}\psi_{n}=E^{\text{osc}}_{n}\,\psi_{n},
\end{equation}
with eigenfunctions 
\begin{equation}
\psi_{n}=\frac{1}{\sqrt{(2\mu\hbar\nu)^{\,n}\,n!}}\,(a^{\dagger})^{n}\,|0\rangle,
\end{equation}
and energy eigenvalues given by
\begin{equation}
E^{\text{osc}}_{n}=\frac{\hbar\nu}{2}\,(2n+1),\quad n\in\mathbb{Z}_{\ge0}.
\end{equation}
as analogous levels of Landau.

On the other hand, the linear part $\hat{K}$ satisfies the follow eigenvalue equation
\begin{equation}
\hat{K}\,\psi_{\alpha}=E^{\text{lin}}_{\alpha}\,\psi_{\alpha},
\end{equation}
with solutions of the form
\begin{subequations}
\begin{eqnarray}
\psi_{\alpha} &=& \exp\!\left(i\alpha\,y-\frac{\mu\nu}{2\hbar}\,x\,y\right);\\
E^{\text{lin}}_{\alpha} &=& \frac{\hbar\lambda}{\mu}\,\alpha+\frac{\lambda^{2}}{2\mu}\,\,\ \text{with}\,\  \alpha\in\mathbb{R}.
\end{eqnarray}
\end{subequations}
The label $\alpha$ accounts for the Landau degeneracy in this gauge; normalization is understood in a finite area (box normalization) or via wave-packet superpositions, as in the usual Landau problem.

Therefore the full eigenstates of $\hat{H}$ are
\begin{equation}
\Psi_{n,\alpha}=\psi_{n}\otimes\psi_{\alpha}\equiv|n,\alpha\rangle,
\end{equation}
with the energy eigenvalues associated,
\begin{equation}
E_{n,\alpha}=\frac{\hbar\nu}{2}\,(2n+1)-\frac{\hbar\lambda}{\mu}\,\alpha-\frac{\lambda^{2}}{2\mu}.
\end{equation}
Using $\lambda=\mu c\,\mathcal{E}/\Lambda$, the spectrum is “Landau + small corrections”: the leading term is \(\hbar\nu\big(n+\tfrac12\big)\), with a linear \(\alpha\)-tilt of order \(\Lambda^{-1}\) and a uniform downward shift of order \(\Lambda^{-2}\). Defining \(\epsilon=\hbar\nu/\Lambda\), in the perturbative regime (\(\epsilon\ll1\)) the tilt is \(O(\epsilon)\) and the shift is \(O(\epsilon^2)\), so the Landau fan remains essentially intact—slightly tilted and slightly lowered. In the intermediate regime, compare \(\epsilon(\hbar\alpha/\mu)\) with \(\hbar\nu\); depending on \(\alpha\), the tilt can affect level crossings and occupations.

%
\section{Concluding remarks}\label{conclusions}
Problems involving noncommutativity are often treated using Moyal products, with observables deformed to preserve covariance. This approach, however, typically makes the resulting operators and equations less transparent. In this work, we adopted a different route: we encoded the noncommutative parameters into the metric so that the geometry itself generates an effective magnetic field. Within this setting, we studied the quantum dynamics of a free particle in a torsional background that induces noncommutativity. Our results show that this choice is appropriate: free particles subject to torsion behave as in the usual Landau problem in flat space, as reflected in the resulting energy spectrum.

Starting from the Lagrangian, we derived the coordinate commutation relations in the presence of a strong (effective) magnetic field. For large values of $\Lambda$ or small mass $\mu$, the system is constrained to remain in the ground state, just as in the presence of a real magnetic field. We then rewrote the Hamiltonian in an algebraic form using creation and annihilation operators to obtain the spectrum and eigenstates. In this formulation, the oscillatory contribution dominates when $\Lambda$ is sufficiently large, consistently reproducing the Landau-like structure with the expected corrections.

\acknowledgments{This work was par\-tial\-ly sup\-por\-ted by PRO\-NEX/FA\-PESQ-PB, CNPq and CA\-PES (PRO\-CAD). G. Q. Garcia would like to thank Fapesq-PB for financial support (Grant BLD-ADT-A2377/2024). The work by C. Furtado is supported by the CNPq (project PQ Grant 1A No. 311781/2021-7)}.
\bibliography{references}  

\end{document}